\pdfoutput=1
\documentclass[
aip,apl
 sd,
 amsmath,amssymb,
]{revtex4-1}

\usepackage{graphicx}

\usepackage[caption=false,position=t]{subfig}
\usepackage{dcolumn}
\usepackage{bm}
\usepackage{natbib}
\usepackage{microtype}
\usepackage{comment}

\begin{document}


\title{Efficient, high-speed ablation of soft tissue with few-microjoule, femtosecond pulse bursts}

\author{Can Kerse}
\affiliation{Department of Electrical and Electronics Engineering, Bilkent University, Ankara, 06800, Turkey}

\author{Seydi Yavas}
\affiliation{Institute of Materials Science and Nanotechnology, Bilkent University, Ankara, 06800, Turkey}
\affiliation{FiberLAST, Inc., Ankara, 06531, Turkey}

\author{Hamit Kalayc{\i}o\u{g}lu}
\affiliation{Department of Physics, Bilkent University, Ankara, 06800, Turkey}

\author{Mehmet D. A\c{s}{\i}k}
\affiliation{Nanotechnology and Nanomedicine Department, Institute of Science, Hacettepe University, Ankara, 06800, Turkey}

\author{\"{O}nder Ak\c{c}aalan}
\affiliation{Department of Physics, Bilkent University, Ankara, 06800, Turkey}
\affiliation{Department of Electrical and Electronics Engineering, TOBB University of Economics and Technology, Ankara, 06530, Turkey}

\author{F. \"{O}mer Ilday}
\affiliation{Department of Electrical and Electronics Engineering, Bilkent University, Ankara, 06800, Turkey}
\affiliation{Department of Physics, Bilkent University, Ankara, 06800, Turkey}

\date{\today}



\begin{abstract}
Femtosecond pulses hold great promise for high-precision tissue removal. However, ablation rates are severely limited by the need to keep average laser power low to avoid collateral damage due to heat accumulation. Furthermore, previously reported pulse energies preclude delivery in flexible fibers, hindering {\em in vivo} operation. Both of these problems can be addressed through use of groups of high-repetition-rate pulses, or bursts. Here, we report a novel fiber laser and demonstrate ultrafast burst-mode ablation of brain tissue at rates approaching 1 mm$^3$/min, an order of magnitude improvement over previous reports. Burst mode operation is shown to be superior in terms of energy required and avoidance of thermal effects, compared to uniform repetition rates. These results can pave the way to {\em in vivo} operation at medically relevant speeds, delivered via flexible fibers to surgically hard-to-reach targets, or with simultaneous magnetic resonance imaging.
\end{abstract}

\maketitle




\subsection{Introduction}
Surgical potential of the laser was recognized immediately following its invention \cite{Solon1961}. Within a year after the demonstration of the first laser, an opthalmic application has been shown, in which a detached human retina was restored \cite{CAMPBELL1963}. While other medical applications of lasers in Continuous Wave (CW) regime such as Laser Interstitial Thermal Therapy (LITT); or in quasi CW and nanosecond to microsecond pulsed regime such as urological applications have proliferated, medical operations in which tissue is removed precisely with a laser are limited mostly to ophthalmology. This is in part due to the difficulty of preventing collateral damage to surrounding tissue during irradiation. This issue is resolved, in principle, by ablation using femtosecond pulses with low average power, typically at low (Hz or kHz) repetition rates \cite{Mazur2009}. However, this method suffers from a very low rate of ablation. While this is tolerable in ophthalmology due to the small volumes needed to be ablated \cite{niemz2007laser}, it renders most tissue removal applications of lasers, including hard tissue ablation uncompetitive compared to mechanical removal techniques \cite{schelle2013ultrashort}. A straightforward way to increase the removal rate is to increase the laser repetition rate. However, the average power is also increased, resulting in significant collateral tissue damage due to accumulated heat effects. 

\begin{figure}[h!]
\includegraphics[width=\textwidth]{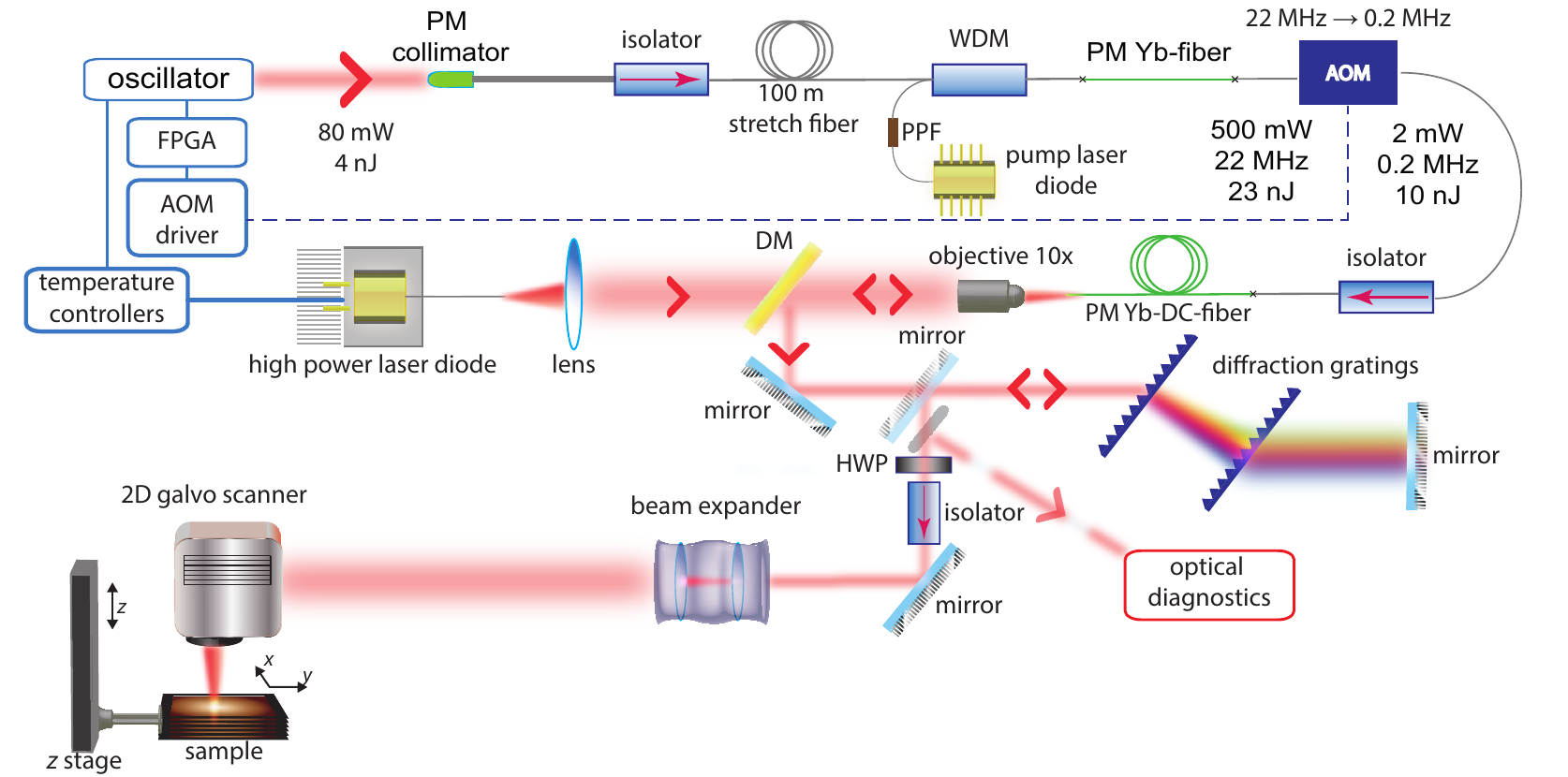}
\caption{\label{setup} Schematic diagram of the laser and tissue processing setups; PM, polarization maintaining; PPF, pump protection filter; WDM, wavelength division multiplexer; DC, double clad; DM, Dichroic Mirror; HWP, half-wave plate.}
\end{figure}

In terms of underlying laser technology, ultrafast tissue ablation experiments have traditionally relied on solid state lasers, particularly Ti:sapphire lasers, which typically produce tens of $\mu$J to mJ levels of pulse energies at kHz or lower repetition rates \cite{fischer1994,juhasz1999corneal,giguere2007laser,palanker2010,hoy2014clinical}. Meanwhile, medical applications of fiber lasers are emerging with major practical advantages, such as compact size, lower cost, superior robustness against environmental fluctuations that are highly desirable in a medical settings. Recently, a new operating mode of fiber lasers have been demonstrated, i.e., the burst mode operation\cite{kalaycioglu2011fiber,kalayciouglu20121,Jena12}. In ultrafast burst-mode operation, which was invented by R. Marjoribanks, et al. \cite{lapczyna1999ultra}, bursts of pulses are generated, each burst consisting of a number of temporally closely spaced (order of 10 ns) pulses. Through the use of ultrafast pulses delivered in burst mode, a completely new interaction regime opens up that holds great potential for use of the laser as a precise scalpel for tissue removal applications, and the possibility to overcome the collateral damage and ablation rate challenges simultaneously \cite{marjoribanks2012ablation,qian2014pulsetrain}. This mode of operation has already been shown to increase the ablation rate for metals \cite{hu2010modeling,kerse2013non} and hard tissue, {\em e.g.} tooth samples \cite{marjoribanks2012ablation, kerse2013non}, as compared to the application of pulses at a lower repetition rate. 

\begin{figure}
\centering
\includegraphics[width=\textwidth]{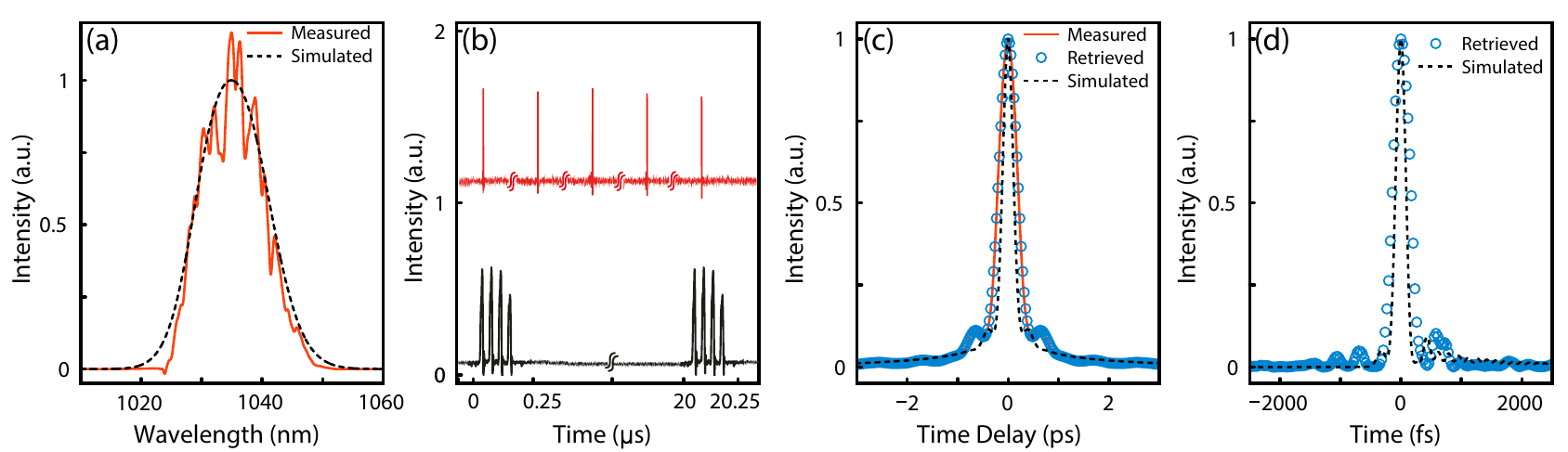}
\caption{\label{IRimage}For 8 $\mu$J compressed pulse energy: (a) measured optical spectrum, (b) measured temporal profile of 200 kHz pulses and 4-pulsed bursts, (c) measured, simulated and PICASO-retrieved autocorrelation,  (d) inferred pulse shape using the PICASO algorithm and simulated pulse shape.}
\end{figure}

Here, we report on a novel Yb-doped fiber laser operating in the burst mode, as well as uniform repetition rate and assess the ablation performance of burst mode on soft-tissue from a medical perspective. The laser can deliver pulses with energy of 8 $\mu$J and duration of $\sim$300 fs at 1030 nm. This constitutes the shortest burst-mode pulses with fiber lasers, to our knowledge. We show that burst mode operation is superior in terms of energy required, ablation speed and avoidance of thermal effects, compared to uniform repetition rates. We demonstrate heat damage-free ablation of rat brain tissue at a rate of 0.75 mm$^3$/min, representing an order of magnitude improvement on previous soft-tissue ablation reports using low-repetition-rate, high-energy solid state lasers. To our knowledge, this is the first time soft-tissue is ablated with a burst mode laser. Notably, the histological studies of the burst-mode ablated areas indicate no collateral damage. Successful ablation of soft tissue using as low as few-microjoule pulse energies in the burst mode opens up the exciting possibility of using hollow-core fibers to deliver ultrafast pulses to remote parts of the body \cite{clark2001fiber,lanin2012air}. With this fiber mode of delivery coupled with the substantial increase in ablation rate over previous reports, our results have potential to render ultrafast tissue processing competitive for an entirely new set of medical operations \cite{Mazur2009}.

\subsection{Materials and Methods}
The experimental setup is depicted in Fig. 1. The seed pulses originate from a Yb-doped fiber oscillator operating in the all-normal dispersion regime at a repetition rate of 22.3 MHz. Mode-locking is initiated and stabilized by nonlinear polarization evolution. The oscillator generates 4-nJ (80 mW) pulses centered at 1037 nm with a spectral width of 15 nm (Fig. 2 (a)) and seeds a 100 m-long polarization-maintaining (PM) fiber stretcher. The seed pulses are thus stretched to 55 ps. This is followed by an in-line fiber pre-amplifier (Yb1200-6/125DC-PM, Liekki), which boosts the signal level to 500 mW. The repetition rate is either reduced to a uniform value of 200 kHz or bursts of pulses are produced at 50 kHz (Fig. 2 (b)) in a fiber-coupled acousto-optic modulator (AOM). After the AOM, the power is reduced to 2 mW. The AOM is controlled by a special FPGA-based (SPARTAN 3E, Xilinx) electronic circuitry developed in house for synchronous pulse picking, burst formation and acts as a master controller for the entire system.

At the power amplifier stage, the setup consists of a 10-W pump diode stabilized to 976 nm by temperature control, a short section of double-clad highly doped Yb gain fiber with core diameter of 20 $\mu$m, cladding of 125 $\mu$m and optics for backward pumping (Yb1200-20/125DC-PM, Liekki). This configuration minimizes the nonlinear effects. Nearly 75\% of the pump power is coupled into the gain fiber with a collimating lens of 25 mm focal length and a 10x objective. Amplified signal is extracted via a dichroic mirror which transmits the pump and reflects the signal wavelengths. We used numerical simulations to guide the experimental design. The details of the simulations can be found in  \cite{oktem2010soliton}. The simulations predict pulse duration of 200 fs for 8 $\mu$J shown in Fig. 2 (c). The retrieved pulse form (Fig. 2 (d)) from measured autocorrelation and spectrum with PICASO algorithm \cite{picaso1999}, indicates $\sim$300 fs FWHM. This result is in close agreement with the simulation, while the effective pulse width of 375 fs of the retrieved pulse reflects the effect of the pedestal caused by the interaction of uncompensated third order dispersion and nonlinearity. Even higher pulse energies (up to 40 $\mu$J) can be generated, but result in longer pulses due to increased nonlinear phase accumulation and are not considered here.

Following dechirping, a high-power isolator is placed to prevent back reflections from the target. The optimized transmission of the compressor, half-wave plate and polarizing isolator (which allows continuous reduction of power) is 80\%. The beam is then expanded with a telescope to completely fill the aperture of a 2D galvanometer scanner (ScanCube 14, ScanLab). The scanner focuses the beam with an f-theta objective, which ensures a constant spot size within a plane. The spot diameter is estimated to be $\sim8$ $\mu$m, which is confirmed by directly observing craters formed after a raster scan on the surface of a copper plate via reflection microscope and independently by the knife-edge method. The target sample is positioned on a motorized 3D stage, which allows translation by 10 cm in 3 axes.

The samples used in the experiments are agar (Agar No: 1, LabM) hydrogel blocks (2 g of agar per 100 ml of distilled water) and freshly harvested rat brain tissues, which were processed within an hour after dissection. In these experiments, we tested the effect of two operating modes, {\em i.e.}, 200 kHz uniform repetition rate mode, and 4 pulsed-burst mode with burst repetition rate of 50 kHz and intra-burst pulse repetition rate of 22.3 MHz.  Hydrogel is preferred especially to compare two operating modes because it is easy to mold to a flat surface, easy to characterize under light microscopy and is a naturally derived hydrogel polymer that has similar molecular properties to extracellular fluid in biological tissues \cite{drury2003hydrogels}. In all of the mode comparison experiments, the pulse duration, delivered fluence and average power are all kept constant. 

For the hydrogel experiments to ensure that the focal plane coincides with the target surface, calibration lines were processed at a speed of 5 mm/s prior to ablation. The optimum focal point was determined by maximizing the brightness of the plasma generated on the surface. The scan speeds were adjusted to 125 mm/s such that approximately 8 pulses for uniform mode and 2 bursts (4 pulses/burst) for burst mode were incident per focal spot. Fig. 3 (a) shows the light microscope image of the hydrogel lesions, created at different individual pulse energy values and compared for burst and uniform repetition rate modes. 

\begin{figure}
\centering
\includegraphics[width=\textwidth]{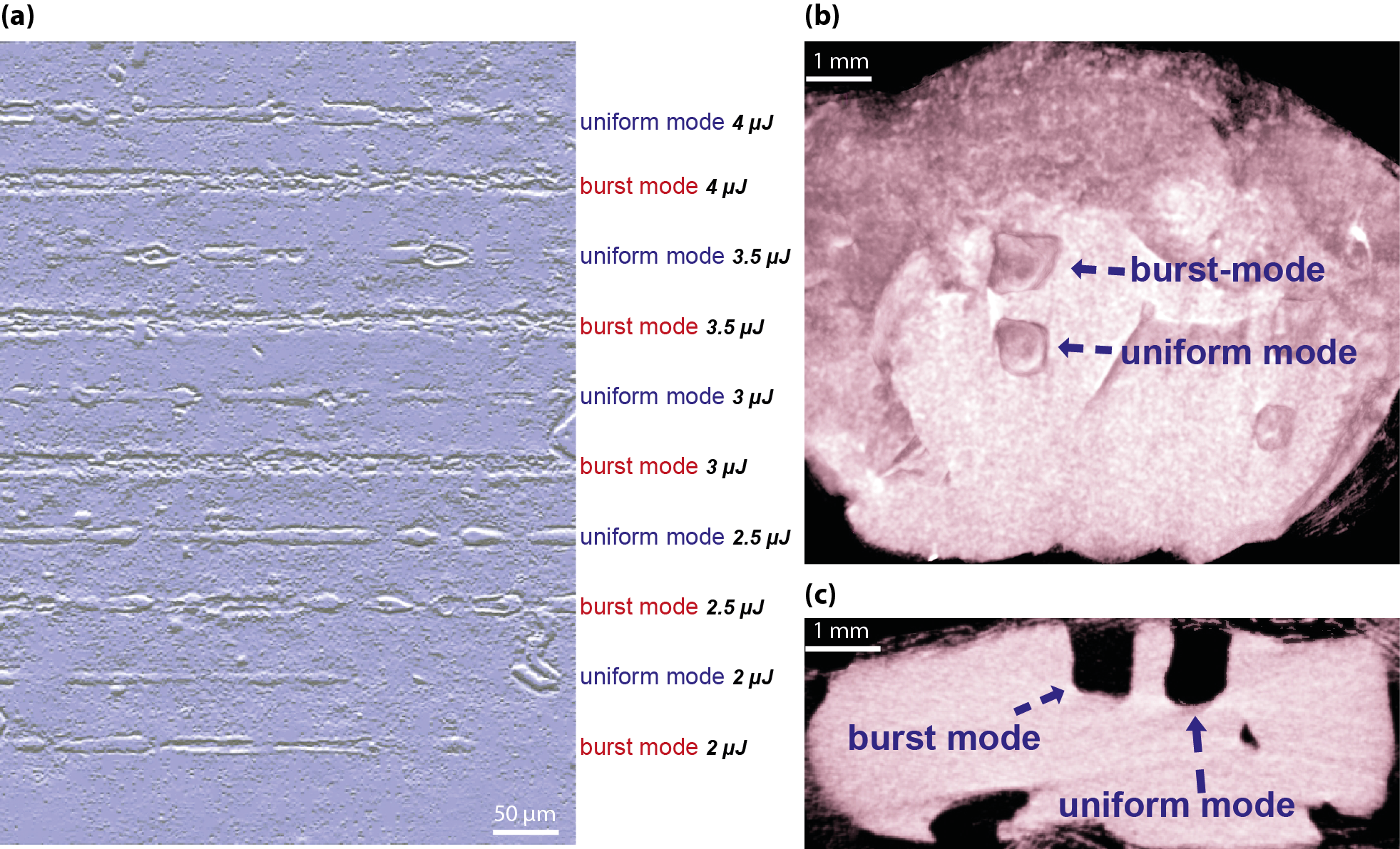}
\caption{(a) Microscope image of hydrogel block, raster-scanned with 200-kHz uniform repetition rate and 4-pulsed burst mode at the indicated pulse energies. Micro CT images of the processed rat brain tissue sample: (b) Top view of the coronal slice where four excisions made in (1) 4 pulsed-burst mode and (2) 200-kHz uniform repetition rate. (c) Sagittal view of the excisions in (b).}
\label{fig:hydrogel and tissue}
\end{figure}

For the tissue experiments, following extraction, the brains were sliced parallel to the coronal plane and the slice to be used was immediately placed on the processing platform. The pulse energy was set to 3 $\mu$J at an average power of 600 mW in the experiments. We performed raster scans over an area of 1 mm x 1 mm with line separation of 2.5 $\mu$m and scan speed of 250 mm/s and with 80\% duty cycle to avoid excessive heating. The raster scan was then repeated 50 times with total exposure time of 80 s. During the experiment, NaCl solution was sprayed regularly on the samples to prevent the tissue from drying. At the beginning of the experiment and in between consecutive raster passes, the optimum focal spot was readjusted to match the surface based on brightness of the plasma generation. After irradiation, the samples were preserved in 4\% paraformaldehyde solution, and then examined with a micro-CT scanner (Skyscan 1172, Bruker). Micro-CT analysis was made with 60 kV and 160 $\mu$A with 0.5 mm Al filter and at a resolution of 12 $\mu$m/pixel. Fig. 3(b) and (c) show the images of processed tissue. Histological analyses were also made on several samples. Following fixation, the tissue samples were embedded into paraffin blocks and then cut into 5 $\mu$m slices in sagittal plane, perpendicular to tissue surface. Then, slices were examined under light microscopy (Leica, DM 5000 B) following a standard hemotoxylin eosin staining procedure. 

\subsection{Results and Discussion}

\begin{figure}
\centering
\includegraphics[width=\linewidth]{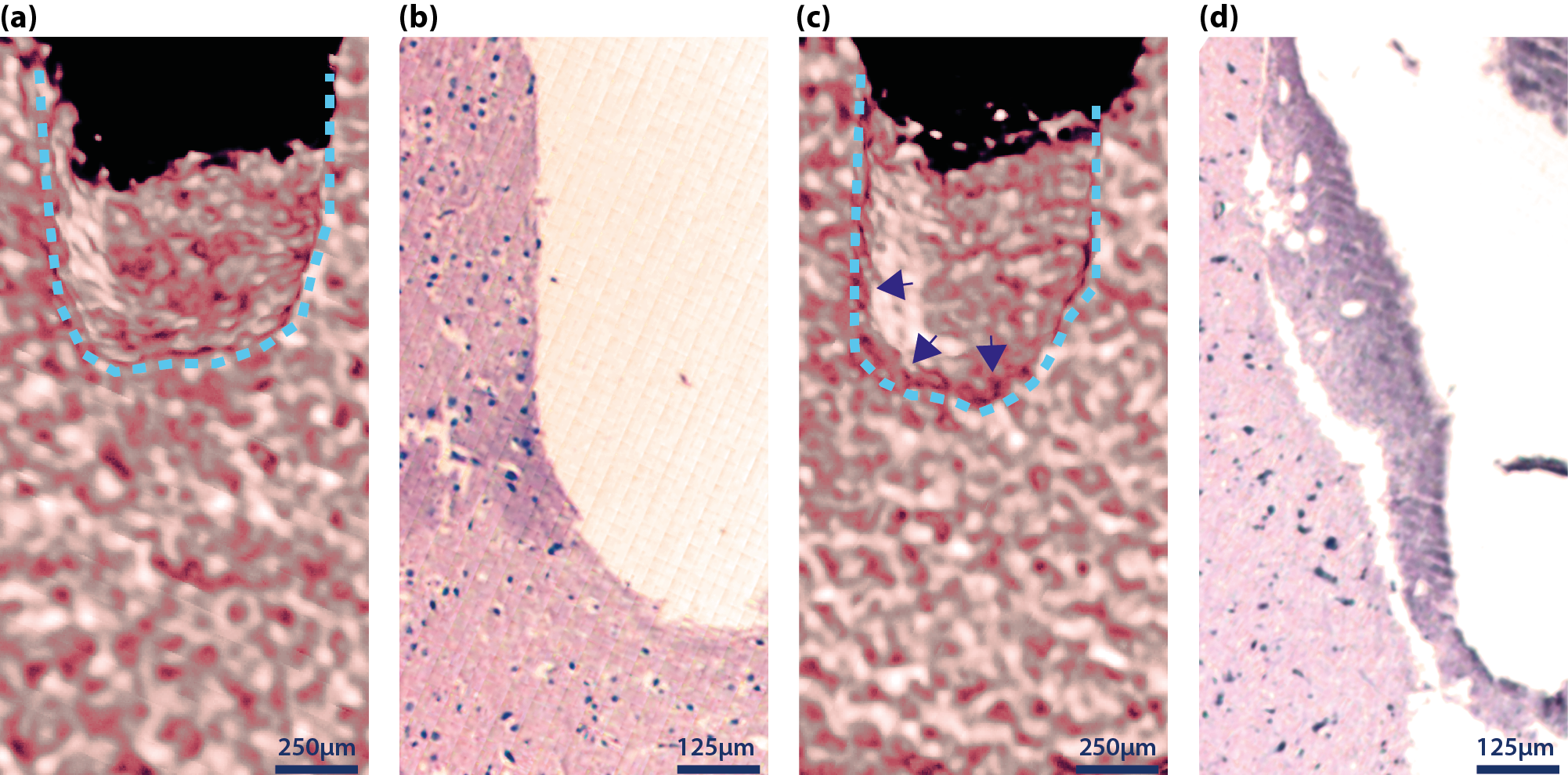}
\caption{(a) Color and contrast enhanced sagittal view of 4-pulsed burst ablation region from 3D volumetric reconstruction of Micro CT data. Dashed lines represent the edges of processed region. (b) Light microscopy image of 4-pulsed burst ablation region after histological preparation. (c) Color and contrast enhanced sagittal view of 200 kHz ablation region from 3D volumetric reconstruction of Micro CT data. Dashed lines represent the edges of processed region. Arrows show the thermal damage zone. (d) Light microscopy image of 200 kHz ablation region after histological preparation.}
\label{fig:histology}
\end{figure}

For hydrogel experiments, it is seen that, at pulse energies exceeding 2.5 $\mu$J, the lesions formed by burst-mode operation are continuous and uniform, whereas the lesions formed by uniform repetition rate mode are not ( Fig. 3 (a) ). Given that the pulse duration, delivered fluence and average power are all identical, these results demonstrate that the cumulative effect of the pulses in a burst leads to increased ablation effectiveness. Using uniform repetition rate pulses, even at 4-5 $\mu$J per pulse, the cuts are not continuous. Utilizing higher pulse energies results in localized melting of the hydrogel sample due to correspondingly higher average power, in addition to precluding delivery through fiber. Reducing the repetition is naturally a possibility, but also reduces the processing speed.

The depth of the lesions on brain tissue were measured to be $\sim$1200 $\mu$m and $\sim$1100 $\mu$m for 4-pulsed burst and 200 kHz uniform repetition rate modes, respectively. Our results indicate that it is possible to ablate $\sim$1 mm$^{3}$ of tissue within 80 s (0.75 mm$^3$/min) without any apparent heat damage to surrounding tissue. In contrast, earlier reports with solid state lasers utilized much higher energies and lower repetition rates. Using 165-$\mu$J, 180-fs pulses at 100 Hz, 0.55-mm$^{3}$ section of bovine brain was ablated in 360 s (0.09 mm$^3$/min) \cite{loesel1998non}. 
No clear difference is observed between uniform repetition rate and burst-mode operation during multiple-pass processing of rat brain tissue. This is likely due to the fact that ablation process halts after several scans over a given point and is re-initiated after re-adjustment of the focal plane, which was done manually in our experiments. The ablation speed can likely be improved substantially with an automated focusing system. 

Even if the ablation depths are quite similar for both 4-pulsed burst mode and 200 kHz uniform repetition rate mode (Fig. 3 (c)), histological analysis reveals that there exists tremendous amount of heat affected zone for 200 kHz mode ablation, as compared to 4-pulsed burst mode. 4-pulsed burst application has minimized collateral damage to the tissue. On the other hand, 200 kHz application damaged cells. Also devascularization can be seen in Fig. 4 (d). Very little amount of tissue damage is observed in 4-pulsed burst application, which is localized to a specific area as seen in Fig. 4 (b). This might be due to the ablation scheme and improper focus adjustment. However, there is a prominent tissue loss and damage under 200 kHz application site. Fig. 4 (a) and (c) shows the zoomed in side views of ablation craters  of 3D reconstructed volumetric images of the regions processed with 4-pulsed burst and 200 kHz modes, respectively. The dashed lines show the edges of the ablation. The images were enhanced in terms of contrast, brightness and saturation in order to emphasize on the tonality difference between normal tissue and thermal damage region in Fig. 4 (c). The darker color in between the dashed lines and the arrows represents the heat affected zone and is in good agreement with the histological cross section of Fig. 4 (d). This contrast might be due to 1-2\% x-ray attenuation difference between normal tissue and the tissue with edema \cite{phelps1975attenuation}. 

The histological differences between the two operating modes are due to repetition rate differences and interaction regime of the laser pulses with the tissue, as reflected in different ablation efficiencies. In our experiments for both modes of operations, the scan speed was adjusted so that four pulses were delivered per spot. This corresponds to delivery of that four pulses in 200 ns and an idle period of 20 $\mu$s for burst mode and in 20 $\mu$s with the period of 5 $\mu$s for 200 kHz uniform repetition rate mode. The thermal relaxation time of biological tissues for 100 $\mu$m spot diameter and with linear absorption coefficient of 1000 cm$^{-1}$ is in the order of 100 $\mu$s for a pulse duration of 1$\mu$s \cite{choi2001analysis}. However, in our case, the interaction regime is totally different, therefore, linear absorption has very little effect on ablation, whereas, absorbed energy density is expected to be trapped in a much more shallower region \cite{oraevsky1996plasma}. Besides, during a burst, incoming pulse is expected to be delivered on denser plasma at a higher temperature because of 50 ns time separation between consecutive pulses, which will increase the efficacy of that pulse and cause quasi-continuous material ejection. On the other hand, for 200 kHz uniform repetition rate mode, since the separation between pulses is 5 $\mu$s, the effect of the incoming pulse will be much less and the residual heat caused by the previous pulse would already be transferred to the deeper regions. In other words, incoming pulse spends most of its energy to re-heating of the surface rather than delivering it to the plasma. Considering these, the relaxation time could be between 5 $\mu$s and 20 $\mu$s.

Our results were achieved with a low pulse energy of 3 $\mu$J, which paves the way toward photonic crystal fiber-delivery of the pulses to which will enable to reach remote parts of the body. Fiber delivery can further enable combination of tissue processing with simultaneous magnetic resonance imaging (MRI) to be used for precise tissue removal or functional MRI studies. Another advantage of lower pulse energies is to minimize cavitation bubble formation \cite{juhasz1999corneal}. 


\subsection{Conclusion}
In conclusion, we report the first successful application of burst-mode femtosecond pulses to soft tissue using a unique fiber laser. The custom-developed fiber laser can operate either in uniform repetition rate pulse or ultrafast burst-mode regimes. We have then applied this laser to ablation of rat brain tissue, achieving ablation rates of 0.75 mm$^3$/min using the burst mode. Histological analysis confirmed that there was no discernible heat-induced damage in the surrounding area of the processed tissue. In contrast, histological analysis shows that the uniform repetition rate of operation of the laser under otherwise identical conditions resulted in substantial heat-induced damage. The burst-mode results represent nearly an order of magnitude improvement on previously reported ablation rates of brain tissue using solid-state lasers \cite{loesel1998non}. The fiber laser parameters and burst-mode operation explored in this work appear to be well-suited to a variety of medical applications including neurosurgery and ophthalmology \cite{palanker2010,juhasz1999corneal,tsai2009plasma}. Applications of the present laser system to femtosecond cataract surgery is currently underway in our laboratory; preliminary results indicate that the required pulse energy is reduced for the burst mode, which is expected to reduce complications \cite{marjoribanks2012ablation,fs_sideeffects,cataract}. While the laser system is capable of generating 8-$\mu$J, 300-fs pulses in both regimes, high-rate and non-thermal ablation could be achieved with pulse energies as low as 3 $\mu$J in the burst mode operation. This energy level is low enough that delivery within a flexible, singlemode, hollow-core fiber becomes a possibility as demonstrated recently for uniform repetition rate pulses in Ref. \cite{lanin2012air}. Many exciting future developments, which can benefit from ultrafast burst-mode laser technology, can be envisioned considering possibility of delivery via hollow-core fibers, paving the way for delivery to otherwise inaccessible areas inside hollow-core fibers. Fiber delivery from a distant laser would allow operation inside magnetic resonance imaging (MRI) systems, opening up entirely new vistas for MRI-guided delivery and processing of soft tissue at medically relevant speeds, which is to date restricted to laser-induced thermotherapy \cite{MRI_2010}. Recently, deep brain imaging via three photon microscopy requires pulse energies approaching 100 nJ \cite{Chris_Xu}. However, live brain imaging requires multi-MHz repetition rates to avoid signal distortion due to, e.g., blood flow. High repetition rates within the burst mode would allow high-speed signal acquisition, while average power can be kept arbitrarily low by adjustment of the duty cycle. Various other biomedical applications can benefit similarly, ranging from two-photon optogenetics \cite{optogenetics, 2Poptogenetics}, to even {\em in-vivo} bioscaffold fabrication \cite{scaffolds, 2Pscaffolds}.


\subsection*{Funding Information}
This work was supported partially by T\"{U}BITAK under projects no. 112T980 and no. TEYDEB-3110216 and the European Research Council (ERC) Consolidator Grant ERC-617521 NLL. 

\subsection*{Acknowledgements}
The authors thank Gamze Aykut for animal care and brain slicing, Arda B\"{u}y\"{u}ksungur (Ph.D.) and BIOMATEN (ODT\"{U}, Ankara, Turkey) for the micro-CT analysis and Onur Tokel (Ph.D.), Haydar Celik (Ph.D.) and Oktay Algin (M.D.) for critical reading of the manuscript. 

\bibliographystyle{naturemag}
\bibliography{Kersebibliography}

\end{document}